\documentclass[nohyper,12pt,letterpaper]{JHEP}
\usepackage{epsfig}

\newfont{\frak}{eufm10 scaled 1200}

\newfont{\Bbb}{msbm10 scaled 1200}     
\newcommand{\mathbb}[1]{\mbox{\Bbb #1}}
\DeclareSymbolFont{AMSa}{U}{msa}{m}{n}
\DeclareSymbolFont{AMSb}{U}{msb}{m}{n}
\let\Box\relax
\DeclareMathSymbol{\Box}{\mathord}{AMSa}{"03}

\title{The Holographic Approach to Cosmology}
\author{T. Banks${}^*$\\
   Department of Physics and Astronomy - NHETC\\
   Piscataway, NJ 08540\\
   and\\
   Department of Physics, SCIPP\\
   University of California, Santa Cruz, CA 95064\\
E-mail: \email{banks@physics.rutgers.edu}}

\author{W.Fischler${}^{**}$\\
Department of Physics and Astronomy\\ University of Texas, Austin,
TX
\\ E-mail: \email{fischler@physics.utexas.edu}}

\abstract{We review the successes and challenges of the holographic
approach to cosmology.  The model predicts an {\it exactly} scale
invariant fluctuation spectrum with long and short distance
cut-offs.  It can account for the observed fluctuations in the CMB
and might explain the low power at large scales. We outline various
cosmological histories compatible with holographic initial
conditions. This paper is based on talks given by the authors at
Cosmo 04 in Toronto, and the 2004 Tamura Symposium in Austin}


\preprint{\hepth{0412097}\\SCIPP-04/21\\UTTG-09-04}

\begin{document}




\section{\bf Introduction}


Holographic Cosmology\cite{tbf1234} is a quantum mechanical approach
to cosmological initial conditions.   The covariant entropy bound
\cite{fsb} bounds the entropy in causal diamonds for early time
observers in FRW cosmology by a function of cosmic time, which
decreases as one approaches the Big Bang. The entropy of any density
matrix is bounded by the logarithm of the dimension of the quantum
Hilbert space, and we have conjectured that the covariant entropy
bound should be interpreted in terms of this dimension.   If we
accept the postulate that the description of a causal diamond does
not depend on physics outside the diamond, this bound shows that no
conventional field or string theoretic description can describe
observations in the early universe, because they have too many
degrees of freedom.

We have, instead, provided an explicit quantum model of early causal
diamonds in terms of a sequence of Hamiltonians drawn from a certain
random ensemble of Hamiltonians in finite dimensional Hilbert
spaces\cite{bfm}.  We showed that this system had the following
properties

\begin{itemize}

\item It solved a set of plausible consistency conditions\cite{bfm} for any
quantum description of a generally covariant system.

\item In the limit when the Hilbert space dimension was large, it
obeyed all of the scaling laws of Friedmann Robertson Walker (FRW)
cosmology with equation of state $p=\rho$ and flat spatial sections.
The equation of state, and flatness, follow from the quantum
dynamics, and are exact properties of the model. In addition, the
full quantum dynamics had an exact symmetry under the conformal
Killing transformation of the $p=\rho$ geometry.

\item This enables one to prove a scaling law for fluctuations
around the homogeneous $p=\rho$ background.  We will review this
below.

\item  All the features of our heuristic description of the $p=\rho$
cosmology as a dense black hole fluid\cite{holocosm1} were exhibited
as properties of the mathematical model.

\item All of these features follow for generic initial conditions in
the model.
\end{itemize}

Holographic cosmology is based on the assumption that our own
universe began as a close relative of the model cosmology of
\cite{bfm}.   That is, it was a somewhat less entropic choice of the
initial conditions.   We picture it as a more or less normal FRW
region embedded as a subset of the coordinate space of the $p=\rho$
universe. Here, as always, we work in coordinates where the causal
diamond reaching from each point on a time slice, back to the Big
Bang, has the same area.  We can think of this in terms of phases of
a black hole fluid. In conventional cosmology a dilute black hole
fluid has equation of state $p=0$. Our model provides evidence for
another phase, with equation of state $p=\rho$, which has larger
entropy density.   Our universe begins as an amorphous region of the
$p=0$ phase, embedded in the $p=\rho$ phase.  We call this region
the {\it cosmic emulsion}.

The Israel junction condition gives us some idea of the constraints
on this region.  We first apply it to a sphere.  Matching the
geometry, on equal area time slices, we find that the coordinate
size of a $p = w \rho $ ($w < 1$) sphere must shrink\footnote{de
Sitter space is an exception.  We can match the horizon of a causal
patch to the horizon of a black hole in the $p=\rho$ universe, with
no coordinate shrinkage of the dS region. We return to this in the
conclusions}. We conclude that the initial region must be a
complicated ``cosmic emulsion" made of many connected spheres. Let
$\epsilon$ be the minimal ratio of normal volume to dense black hole
fluid volume, consistent with survival of the normal region.   We
find that after a cosmic time $T \sim {1\over \epsilon^2}$, the
volume of the universe (or of the largest coordinate sphere
surrounding the cosmic emulsion) is dominated by a normal radiation
gas. The dilute black hole gas, originally occupying the normal
region, quickly decays to radiation by the Hawking process.

The regions of dense black hole fluid in the interstices of the
cosmic emulsion, now behave like black holes of average size $T$,
embedded in a radiation dominated universe.   This black hole fluid
is on the edge of the transition between the two phases.  The
physical volumes covered by black holes and by radiation gas are
approximately equal. If we wait one ten-folding of the universe the
dilute phase is established.  The large black holes now dominate the
energy density of the universe but they behave as a dilute gas.   It
is only at this point in time that conventional notions of field
theory in a background space-time begin to make sense.  Fluctuations
in the energy density which were generated during the $p = \rho$ era
are now encoded in fluctuations in the number density and masses of
black holes.   For consistency of the picture, the fluctuations must
be small.  Order one fluctuations would lead to larger black holes,
via collision processes, and the universe would quickly evolve back
into the dense black hole fluid phase.  We do not yet have an
estimate of how small the fluctuations must be in order for a normal
region to survive.

The energy density (we always use Planck units) at the point when
normal field theoretic cosmology becomes valid is ${1\over a^3 T^2}$,
where we have estimated $a \sim 10$.   Our model assumes that there
is a field\footnote{The single field could be an effective
description of a single trajectory in a multi-dimensional field
space.} whose potential energy density has the form
$$\mu^4 v({\phi \over m_P})$$, where $m_P$ is the reduced Planck mass.
We further assume that $\mu^4 \leq (10)^{-3} T^{-2}$, so that as the
scale factor $a$ increases, the black hole energy density becomes
smaller than the energy density in this field.   Note that since the
holographic initial conditions have already led to a highly
homogeneous universe, it is entirely plausible that there are
regions where the field $\phi$ is spatially homogeneous over scales
of order ${{m_P} \over {\mu^2}}$.   Furthermore, since the field
evolution begins at energy densities $\gg \mu^4$, its initial motion
is certainly friction dominated.  Thus, the initial conditions are
set up for slow roll inflation.

The fluctuations in black hole density, which were generated during
the $p=\rho$ era, are imprinted on the scalar field, because the
criterion for the beginning of inflation in some region of space is
that the black hole energy density in that region drop below $\mu^4$.
These fluctuations will thus appear in the density of radiation
produced by the decay of the inflaton at the end of inflation.  In
addition, there may be fluctuations produced by the quantum
mechanics of the scalar field during inflation.  Note however, that
the inflationary fluctuations might be quite small if $\mu$ is very
small.

\section{Generation of fluctuations in the $p=\rho$ era}

The mathematical model of \cite{bfm} is very far from quantum field
theory in a background geometry.   The geometry is generated
dynamically, in such a way that the system's entropy automatically
scales like the area.   Thus, we cannot study fluctuations in this
system by using the conventional perturbation theory of Einstein's
equations.   A rigorous study of perturbations would require us to
define another solution of the consistency conditions of \cite{bfm}
corresponding to a small perturbation of the $p=\rho$ solution. This
is hard, and we have not yet done it.

However, the form of the two point fluctuation is completely
determined by the symmetries in the model.  The quantum system
constructed in \cite{bfm} led naturally to the introduction of a
homogeneous isotropic FRW space-time manifold. Imagine that we have
succeeded in modifying the construction, so that we can talk about a
normal region embedded in the $p=\rho$ background, and let $P({\bf
x},t)$ be the density of normal regions in the vicinity of the
space-time point $({\bf x},t)$ .    The model had exact quantum
symmetries corresponding to rotation and translation of the ${\bf
x}$ coordinates, as well as to the conformal transformation $ t
\rightarrow b t, \ \ \ {\bf x} \rightarrow b^{- {2\over 3}}{\bf x}$.
The measure $ d^3 x P({\bf x},t)$ should be invariant under these
symmetries.   As a consequence, the two point function for
fluctuations of $P({\bf x}, t)$ must have the form:
$$<P({\bf x} , t) P({\bf y} , s) > = \int d^3 x e^{i{\bf k (x-y)}}
G(t|k|^{3/2},  s|k|^{3/2})  .$$

As noted above, the energy density fluctuations in the normal region
are fluctuations in the density of interstitial black holes in the
cosmic emulsion.   These are related to $P({\bf x}, t)$ by
convolution with a transfer function
$${\delta \rho \over \rho} ({\bf x},t) = \int_{1}^t  ds d^3 y F(|{\bf x -
y}|, t,s) P({\bf y}, s)$$ The transfer function relates the density
of normal regions in the emulsion to the density of interstices
which are filled with $p=\rho$ matter.   It depends on the local
geometry of the emulsion, but it should fall off at large distances
at all times (its Fourier transform should approach a constant for
$|k|$ smaller than a few inverse Planck lengths.).   The fluctuation
spectrum at the instant, $T$, after the transition to the dilute
black hole gas phase is thus:

$$<{\delta \rho \over \rho} (k,T) {\delta \rho \over \rho} (-k,T)> =
|F|^2 \int_1^T dt \int_1^T ds G(tk^{3/2}, sk^{3/2} ),$$ where we
have factored out the constant value of the Fourier transform of
$F$. Rescaling the integration variables, we get
$$<{\delta \rho \over \rho} (k,T) {\delta \rho \over \rho} (-k,T)> =
{{|F|^2}\over k^3} \int_{k^{3/2}}^{Tk^{3/2}} dt
\int_{k^{3/2}}^{Tk^{3/2}} ds G(t, s ).$$ These are scale invariant
fluctuations over a range of $k$ whose physical size at the time of
the transition runs from the inverse horizon size at that time, down
to a few Planck lengths. The fluctuations must be small, but we do
not have a quantitative estimate of how small.  We also do not have
an argument about higher moments of the distribution.

\section{The scale of fluctuations today}

After the end of the dense black hole era, the universe is matter
dominated while the scale factor increases to $a = (T^2
\mu^4)^{-{1\over 3}}$ .  This is followed by $N$ e-folds of
inflation, after which the universe reheats to a temperature $T_{RH}
= \mu^3$.   The correlated fluctuations, which were of physical size
$T$ at the end of the dense black hole era, have size:

$$R_{RH} =T (T^2 \mu^4)^{-{1/3}} e^N \mu^{-4},$$ at the end of reheating.
Their current size is $R_{RH}{T_{RH} \over T_{NOW}}.$  This should
be compared to the current horizon size, which satisfies $R_{NOW}
T_{NOW} \sim 10^{29}$.  If we wish to assert that the fluctuations
we now observe in the CMB, originated in the $p = \rho$ era, we must
require that the correlation length of fluctuations is at least as
large as $R_{NOW}$.  This leads to the inequality

$$T^{1/3} \mu^{-{7/3}} e^N \geq 10^{29}$$

We must also require that the fluctuations extend over $3$ decades
in wavelength (five decades if we want scale invariant fluctuations
at galaxy scales).  This means $T^{2/3} \geq 10^3 (10^5)$.   Note that
something like the latter inequality is probably required for
self-consistency of our picture, independent of the observations.
The description of the microscopic dynamics of \cite{bfm} by a
$p=\rho$ fluid is valid only at scales much larger than the Planck
scale.  Thus, $T$, the horizon size at the end of the $p = \rho$
era, must certainly be much larger than the Planck scale.

Another observational bound comes from requiring that the reheat
temperature be larger than $6$ MeV, so that nucleosynthesis proceeds
according to the standard theory.   This implies $\mu > 10^{- 7}$.

In principle, one should eventually be able to calculate $T$, $\mu$
and $N$ from a microscopic theory.   At this phenomenological stage
we can only discuss scenarios.   The above bounds apply to the
scenario in which the fluctuations generated during the $p = \rho$
era are the ones we observe on the sky.   It is also possible that
the inflationary period generates the observed fluctuations.   Note
however that the {\it a priori} lower bounds on $T$ suggest that
$\mu$ may be quite small.   The inflationary fluctuations of a
simple slow roll model would then be unobservable, and we would have
to turn to a more complicated hybrid model.

The most exciting scenario is the one we have outlined above.  {\it
It predicts an exactly scale invariant adiabatic fluctuation
spectrum over a range of scales which extends from some $R_{max}$
down to $T^{- {2\over 3}} R_{max}$.}  The UV and IR cut-offs on this
spectrum are rather sharp.   If, for some reason, $R_{max}$ were of
order the current horizon scale, this could explain the low angular
momentum anomalies in the CMB.   Although the fluctuations are
definitely small, we have not yet been able to quantify their size,
nor to determine whether they are Gaussian.   These questions depend
on the microscopic dynamics of the cosmic emulsion.

We do not see a mechanism in this scenario, for generating
gravitational waves of wavelength shorter than $T$ prior to the
beginning of inflation.   Thus, observable gravitational wave
backgrounds could only be generated by inflation.  However, since
$\mu < 10^{15}$ GeV, the inflationary gravitational wave amplitude
is too small to be observed in the foreseeable future. Note that
this part of the conclusion is valid, even if the observable scalar
CMB fluctuations come from inflation.  We consider the conclusion
about gravitational waves generated prior to inflation to be
preliminary.

\section{\bf Conclusions}

Holographic cosmology is a quantum mechanical description of the Big
Bang.   It sets up the proper initial conditions for inflation.   It
also solves the homogeneity, isotropy and flatness problems without
recourse to inflation.   Depending on parameters we are currently
unable to calculate, it may produce an observable, non-inflationary
contribution to CMB fluctuations.   This contribution is adiabatic,
and exactly scale invariant, between sharply defined UV and IR
cut-offs.   Preliminary arguments indicate that the primordial
gravitational wave spectrum predicted by this model is below the
level one can expect to observe in the near future.

One final note about how stable and generic our construction is, must
be addressed. The discussion up to this point referred an infinite
$p=\rho$ universe with an infinite cosmic emulsion embedded in it.
The initial coordinate density of cosmic emulsion is $\epsilon$,
which should be taken as small as possible to obtain the most
generic initial conditions compatible with survival of the normal
region.   A much more probable initial condition would be a finite
cosmic emulsion in an infinite $p=\rho$ universe.   This is
potentially unstable, because the Israel junction condition requires
a sphere of normal region to shrink in coordinate size.  A way to
avoid this conclusion is to insist that the normal region asymptote
to a de Sitter space.   The cosmological horizon of de Sitter space
can be joined to a marginally trapped surface in the $p=\rho$
background.  That is, the normal universe is the interior of a $p =
\rho$ black hole!   One way to understand this is that empty de
Sitter space has the same amount of entropy as the ball of dense
black hole fluid which it displaces.  Thus, although the normal
region of the universe begins as a low entropy region, it eventually
evolves to a maximal entropy region. This introduces the
cosmological constant as a new parameter.  It is clear that in terms
of the measure on initial conditions, a universe which will evolve
to have a large cosmological constant is initially more entropic
than a universe which evolves to small $\Lambda$.  That is, in
causal diamonds much smaller than the event horizon, the dS vacuum
has very small entropy.

Thus, we are led to a picture in which the cosmological constant is
a random variable, with an {\it a priori} measure which favors large
$\Lambda$.   Weinberg's bound\cite{wein} now implies that if our
definition of a ``normal" universe includes the requirement that
galaxies can form, then the most probable normal universe in our
model will have $\Lambda \sim Q^3 \rho_0$.   Here $Q$ is the
amplitude of primordial density fluctuations at horizon crossing,
and $\rho_0$ is the dark matter density at the beginning of the
matter dominated era.   In inflationary models of the fluctuations,
$Q$ is independent of $\Lambda$ for $\Lambda \ll \mu^4$, and the
same is true {\it a fortiori} for our $p = \rho$ generated
fluctuations, which arise even earlier than the inflationary era.
$\rho_0$ is typically determined by microscopic physics and will
also be independent of $\Lambda$ for small $\Lambda$.  Assuming
these numbers take on their real world values (in the eventual
Theory of Everything) then these considerations would ``explain" the
value of $\Lambda$ as corresponding to the most probable initial
condition for a universe containing galaxies.

\section{Acknowledgments}

The research of T.B was supported in part by DOE grant number
DE-FG03-92ER40689, the research of W.F. was supported in part by NSF
grant- 0071512.

%



\newpage
\begin{thebibliography}{19}        

\bibitem{tbf1234} T.~Banks, W.~Fischler,
{\it M-theory observables for cosmological spacetimes,}
hep-th/0102077; T.~Banks, W.~Fischler, {\it An Holographic
Cosmology}, hep-th/0111142; T.~Banks, W.~Fischler, {\it Holographic
Cosmology 3.0}, hep-th/0310288.


\bibitem{bfm} T.~Banks, W.~Fischler, L.~Mannelli, {\it Microscopic
Quantum Mechanics of a $p = \rho$ Universe}, hep-th/0408076.


\bibitem{fsb} W.~Fischler, L.~Susskind,
{\it Holography and Cosmology,} hep-th/9806039; R. Bousso, {\it
Holography in general space-times,} JHEP 9906, 028 (1999)
hep-th/9906022



\bibitem{holocosm1}T.~Banks, W.~Fischler, {\it An Holographic
Cosmology}, hep-th/0111142

\bibitem{wein}S.~Weinberg, {\it Anthropic Bound on the Cosmological
Constant}, Phys. Rev. Lett. 59, 2606, (1987).

\end{thebibliography}
\end{document}